\documentclass[11pt]{article}

\usepackage{amsfonts,amssymb,amsmath,amsthm,amscd,latexsym}

\newcommand{\sect}[1]{\setcounter{equation}{0}\section{#1}}

\def\be{\begin{equation}}
\def\ee{\end{equation}}
\def\bea{\begin{eqnarray}}
\def\eea{\end{eqnarray}}

\def\1{\'{\i}}                           
 
\def\RR{\mathbb{R}} 

 \def\aA{{\psi}} 
\def\bB{{\phi}} 
\def\upi{{\varphi}}

\begin{document}

  \hfill\today

\begin{center}
\baselineskip 24 pt {\LARGE \bf  
Drinfel'd doubles for (2+1)-gravity}



\end{center}

\bigskip 
\medskip

\begin{center}

{\sc \'Angel Ballesteros$^1$, Francisco J. Herranz$^1$ and
Catherine Meusburger$^2$}

{$^1$ Departamento de F\1sica, Universidad de Burgos, 
09001 Burgos, Spain}

{$^2$  Department Mathematik,  Friedrich-Alexander Universit\"at  Erlangen-N\"urnberg, Cauerstr.~11, 91058 Erlangen, Germany
}
 
e-mail: {angelb@ubu.es, fjherranz@ubu.es,  catherine.meusburger@math.uni-erlangen.de}

\end{center}

\begin{abstract}

All possible Drinfel'd double structures for the anti-de Sitter Lie algebra ${so} 
(2,2)$  and de Sitter Lie algebra $so(3,1)$ in  (2+1)-dimensions are explicitly constructed and analysed in terms of a  kinematical basis adapted to (2+1)-gravity. Each of these structures  provides in a canonical way a pairing among the  (anti-)de Sitter generators, as well as a specific classical $r$-matrix, and the cosmological constant  is included in them as a deformation parameter.  It is shown that four of these structures  give rise to a Drinfel'd double structure for the Poincar\'e algebra $iso(2,1)$ in the limit where the cosmological constant tends to zero.
We  explain how these Drinfel'd double structures are adapted to (2+1)-gravity,  and we show that the associated quantum groups are natural candidates for the quantum group symmetries  of quantised  (2+1)-gravity models and their associated non-commutative spacetimes.

\end{abstract}
 
\medskip 
\medskip 


\medskip 







\sect{Introduction}

Quantum group symmetries play an important role in  the quantisation of gravity.  They occur both in Hamiltonian quantisation formalisms 
for (2+1)-gravity such as combinatorial quantisation~\cite{AGSI,AS, BR,BNR,we2} and in path integral approaches towards quantum gravity (state sum models, spin foams). In spin foam models,   quantum group symmetries are essential for the construction of the diffeomorphism invariant Hilbert space  and  ensure the convergence of the models. 
In addition to this,  quantum group symmetries  also arise in phenomenological approaches to quantum gravity in three and higher dimensions  \cite{majid88},  in non-commutative geometry models 
such as $\kappa$-Poincar\'e models~\cite{Lukierskia,Lukierskib,Maslanka,Majid:1994cy,LukNR,nullplaneA,Lukierskicc,nullplaneB,Lukierskicd,nullplaneC}  and related  `doubly special relativity'   theories~\cite{Amelino-Camelia:2000mrr,Brunoa,MagueijoSmolin,Kowalski-Glikman:2002we,Brunob,Lukierski:2002df,amel,KowalskiFS}.

While the quantum group symmetries in (3+1)-gravity are often introduced {ad hoc} or on phenomenological grounds, their  emergence in (2+1)-gravity is more transparent, since they arise as the quantum counterparts of  Poisson-Lie symmetries of the classical theory \cite{AMII,FR}.  These Poisson-Lie symmetries emerge from a description of (2+1)-gravity as a Chern-Simons gauge theory \cite{AT, Witten1}, in which  the gauge group  is the isometry group of the corresponding constant curvature spacetimes,  and the phase space of (2+1)-gravity is related to the moduli space of flat connections. 
It is shown in \cite{AMII,FR} that  the Poisson structure on the moduli space has a natural description in terms of Poisson--Lie group and coboundary Lie bialgebra structures associated with the isometry groups. 

The admissible classical $r$-matrices are characterised by the condition that their symmetric component  is dual to the Ad-invariant symmetric bilinear form in the Chern-Simons action.
This relation to Poisson-Lie symmetries on the classical phase space  allows one to draw conclusions about the quantum group symmetries  of (2+1)-gravity by considering their semiclassical limits ---the associated Poisson-Lie groups--- and to clarify their physical and geometrical interpretation.
 In particular, it was shown in \cite{MSquat,MS}
 that  the $\kappa$-Poincar\'e symmetries and their counterparts in (anti-)de Sitter space  are not compatible with the Chern-Simons formulation of (2+1)-gravity.

In this context, the associated quantum groups arise naturally as symmetries of the quantum theory and have a clear physical interpretation. The coproduct determines the composition of observables for multi-particle models as well as the implementation of constraints, while the antipode describes  anti-particles.  The universal $R$-matrix governs the exchange of particles through braid group symmetries, i.~e.~the braiding of their worldlines,  and the ribbon element  the quantum action of the pure mapping class group.

Despite this natural interpretation, the role of quantum group symmetries in (2+1)-gravity is subtle. The first issue is the question of  which quantum deformations of the isometry groups in (2+1)-gravity are suitable for the quantisation of the theory. While $\kappa$-Poincar\'e models have been considered extensively in this context, they are not compatible with the Chern-Simons formulation of (2+1)-gravity as they do not admit a symmetric component which is dual to the Ad-invariant symmetric bilinear form in the Chern-Simons action. Instead, there is evidence that the quantum group symmetries relevant to this context are the Drinfel'd doubles of the (2+1)-dimensional Lorentz group and its Euclidean counterpart~\cite{BNR, we2, MSquat, MS, MajidSch, Mnoui, Woronowicz}. 

Further evidence for the relevance of Drinfel'd doubles in this context arises from the recent  work~\cite{balskir,turvire}  on observables in the context of the Turaev-Viro model. In particular, it was shown  in~\cite{balskir,turvire}  that  the the Turaev-Viro invariant  \cite{TuVir, BWI} only depends  on the center of the underlying spherical category. If the spherical category under consideration is  given as a representation category of a Hopf algebra, then its center corresponds to representations of its Drinfel'd double. As the Turaev-Viro invariant for $U_q(su(2))$  is a state sum model for   Euclidean  (2+1)-gravity with positive cosmological constant, this provides another strong motivation to systematically investigate all Drinfel'd double structures associated with the isometry groups in (2+1)-gravity. 

Another issue related to the role of quantum group symmetries in (2+1) and higher-dimensio\-nal gravity is the role of the cosmological constant $\Lambda$. Most models of quantum gravity  which exhibit  quantum group symmetries  are defined only for a fixed value of $\Lambda$.  This makes it difficult to relate the models for different values of the cosmological constant and to understand  $\Lambda$ as a continuous parameter. It also causes difficulties when  various limits of the models are considered, since the deformation parameter of the quantum group involves a combination of different physical constants.  It was argued in \cite{BHM} that  this could be remedied by considering multi-parametric quantum deformations, in which the different deformation parameters correspond to Planck's constant, the cosmological constant and the speed of light and in which various limits can be realised as quantum group contractions.   Specific examples of deformations of this type were considered in \cite{BHM,bernd1}, but a systematic and complete analysis is still lacking. 

In this article, we analyse systematically all classical  Drinfel'd double structures associated with the Lie algebras $so(3,1)$ and $so(2,2)$, which are the isometry algebras of (2+1)-dimensional de Sitter and anti-de Sitter spaces. We investigate the role of these Drinfel'd double structures in (2+1)-gravity and determine which of them admit a cosmological limit $\Lambda\to 0$,  thus giving rise to a  Drinfel'd double structure on the Lie algebra $iso(2,1)=so(2,1)\ltimes\mathbb R^3$, which is the isometry algebra of (2+1)-dimensional Minkowski space. 
This is achieved by relating all Drinfel'd double structures to a common  basis for these three Lie algebras, in the following referred to as `kinematical basis'. This kinematical basis  has a direct physical interpretation in (2+1)-gravity and involves the cosmological constant  as a structure constant.

As the isometry algebra $so(3,1)$ of  (2+1)-dimensional de Sitter space admits   four different Drinfel'd double structures  and the isometry algebra $so(2,2)$ of  (2+1)-dimensional anti-de Sitter space  admits  three \cite{gomez,Snobl}, this gives rise to seven different cases. For each of them, we determine explicit expressions for the classical $r$-matrix in the kinematical basis and  investigate its cosmological limit  $\Lambda\to 0$. The result is that two of the Drinfel'd double structures for $so(2,2)$ are related to corresponding Drinfel'd double structures for $so(3,1)$ and exhibit a well defined cosmological limit, while the last one diverges when the cosmological constant tends to zero and is not directly related to (2+1)-gravity.  The four Drinfel'd double structures for $so(3,1)$ involve two which are related to Lorentzian (2+1)-gravity with positive cosmological constant and two for Euclidean (2+1)-gravity with negative cosmological constant.

These results can be viewed as a first step towards  the construction of  non-commutative spacetimes, in which both, the cosmological constant  and the Planck constant appear as  deformation parameters. This would require the construction of the full quantum deformations of the (anti-)de Sitter algebras associated with the classical Drinfel'd doubles.  Another avenue for further research is to fully develop the theory of interacting point particles on the semiclassical level by constructing the associated Poisson-Lie structures.


\sect{Spaces and Lie algebras for (2+1)-gravity}

General relativity in (2+1)-dimensions has a simpler structure than its higher-dimensional counterparts.  As the Ricci-tensor of a three-dimensional  manifold determines its curvature, (2+1)-dimensional spacetimes do not exhibit local gravitational degrees of freedom and are locally isometric to certain model spacetimes, which are displayed in Table~\ref{table1}.

  For Lorentzian signature, these are  (2+1)-dimensional   de Sitter space ${\bf dS}^{2+1}$   ($\Lambda>0$),   Minkowski space ${\bf M}^{2+1}$    ($\Lambda=0$) and      anti-de Sitter  space  ${\bf AdS}^{2+1}$  ($\Lambda<0$). 
   The associated isometry  groups are  the  Lorentz group $\mathrm{SO}(3,1)$, the Poincar\'e group $\mathrm{ISO}(2,1)$   and   $ \mathrm{SO}(2,2)= \mathrm{PSL}(2,\mathbb R)\times \mathrm{PSL}(2,\mathbb R)$,  respectively.   For Euclidean  signature, the relevant  model spacetimes are  the three-sphere  ${\bf S}^{3}$  ($\Lambda>0$),   Euclidean space  ${\bf E}^{3}$  ($\Lambda=0$) and    hyperbolic space  ${\bf H}^{3}$  ($\Lambda<0$), whose  isometry groups are, in this order, $ \mathrm{SO}(4)=\mathrm{SO}(3)\times \mathrm{SO}(3) $,  $\mathrm{ISO}(3)$   and  $\mathrm{SO}(3,1)$.
Spacetimes with point particles or non-trivial topology are obtained as quotients of these model spacetimes or  by gluing certain domains in them (for an overview, see \cite{Carlipbook} and the references therein), both of which can be described by   
 group homomorphisms from their fundamental group of the spacetimes  into the isometry groups of the associated model spacetimes \cite{mess, bb}.

An important feature of this description is the fact the   
 Lie algebras ${so}(3,1)$, ${iso}(2,1)=so(2,1)\ltimes \mathbb R^3$,  ${so}(2,2)$  of  the isometry groups  of  Lorentzian (2+1)-gravity and their Euclidean  counterparts 
 $so(4)= {so}(3)\oplus {so}(3)$, ${iso}(3)=so(3)\ltimes \mathbb R^3$, ${so}(3,1)$ can be described in terms of a common basis  $\{J_a,P_a\}_{a=0,1,2}$, such that  the cosmological constant $\Lambda$ plays the role of a structure constant \cite{Witten1}. In this basis, the Lie bracket  takes the form
 \begin{align}
  \label{brack}
  [J_a,J_b]=\epsilon_{abc}J^c , \qquad [J_a,P_b]=\epsilon_{abc}P^c , \qquad [P_a,P_b]=\chi\,\epsilon_{abc}J^c,
  \end{align}  
where, depending on the signature, indices are raised with either the three-dimensional Minkowski metric $g=\text{diag}(-1,1,1)$ or the Euclidean metric $g=\text{diag}(1,1,1)$ and $\chi$ is directly related to the cosmological constant $\Lambda$ through
\begin{align}\label{lambdadef}
\chi=\begin{cases}\Lambda & \text{for Euclidean signature};\\
-\Lambda  & \text{for Lorentzian signature.}\end{cases}
\end{align}
The six Lie algebras arising in (2+1)-gravity with  Euclidean and Lorentzian signature are therefore given by
 \be 
\begin{array}{lll}
 [J_0,J_1]=J_2,  &\quad [J_0,J_2]=-J_1,  &\quad [J_1, J_2]=\alpha\,J_0, \\[2pt]
 [J_0,P_0]=0 ,   &\quad [J_0,P_1]= P_2 ,  &\quad [J_0, P_2]= - P_1,\\[2pt]
[J_1,P_0]=-P_2 ,  &\quad [J_1,P_1]=0 ,  &\quad [J_1, P_2]= \alpha\,P_0,\\[2pt]
[J_2,P_0]=P_1 ,   &\quad [J_2,P_1]= -\alpha\,P_0 ,   &\quad [J_2, P_2]=0,\\[2pt]
[P_0,P_1]=\chi\,  J_2,  &\quad [P_0, P_2]=-\chi\,J_1  , &\quad [P_1,P_2]= \alpha\,\chi\, J_0   , 
 \end{array}
  \label{jj}
\ee 
where  $g=\text{diag}(\alpha,1,1)$ with $\alpha=\pm 1$ denotes the  Euclidean and Minkowski   metric in three dimensions and $\Lambda=\alpha\chi$ as in  \eqref{lambdadef}. 
A direct computation shows that this bracket indeed satisfies the Jacobi identity and hence for all values of the two parameters $\alpha,\chi$ defines a six-dimensional  real Lie algebra.

Note that the basis $\{J_a,P_a\}_{a=0,1,2}$ is distinguished from other bases of this Lie algebra by the fact that its elements  have a direct geometrical interpretation. The basis elements $J_a$ are the infinitesimal generators of Lorentz transformations and rotations for, respectively,  Lorentzian and Euclidean signature,  and the elements $P_a$ generate  translations in the associated (2+1)-dimensional spacetimes.  These translations commute if and only if the curvature of the model spacetime vanishes,  i.~e.~for  $\Lambda=\chi=0$.
As the basis $\{J_a,P_a\}_{a=0,1,2}$ corresponds to the  kinematical  symmetries of (2+1)-gravity,  we will  refer to this basis  as the `kinematical basis' in the following.


\begin{table}[t] {\footnotesize

 \noindent
\caption{{\small Constant curvature spacetimes   and isometry groups  in  (2+1)-gravity in terms of the signature of the metric,  the cosmological constant $\Lambda$ and the values of the parameters $\alpha$, $\chi$.
 }}
\vspace{2ex}
\label{table1}
  \begin{tabular}{llll}
      \hline
 & & & \\[-1.5ex]
 Metric  &  $\Lambda>0$ & $\Lambda=0$ & $\Lambda<0$\\[+1.0ex] 
     \hline
 & & & \\[-1.0ex]
{\footnotesize Lorentzian} & $\alpha=-1$\quad $\chi<0$ & $\alpha=-1$\quad $\chi=0$& $\alpha=-1$\quad $\chi>0$\\[2pt]
   & ${\bf dS}^{2+1}=\mathrm{SO(3,1)/SO(2,1)}$ & ${\bf M}^{2+1}=\mathrm{ ISO(2,1)/SO(2,1)}$ & ${\bf AdS}^{2+1}=\mathrm{SO(2,2)/SO(2,1)}$\\[2pt]
& $\text{Isom}({\bf dS}^{2+1})=\mathrm{SO(3,1)}$ & $\text{Isom}({\bf M}^{2+1})=\mathrm{ ISO(2,1)}$ & $\text{Isom}({\bf AdS}^{2+1})=\mathrm{SO(2,2)}$\\[6pt]
{\footnotesize Euclidean}  & $\alpha=+1$\quad $\chi>0$ & $\alpha=+1$\quad $\chi=0$& $\alpha=+1$\quad $\chi<0$\\[2pt]
  & ${\bf S}^3= \mathrm{ SO(4)/SO(3)}$   & ${\bf E}^3=\mathrm{ ISO(3)/SO(3)}$ & ${\bf H}^3=\mathrm{ SO(3,1)/SO(3)}$\\[2pt]
&$\text{Isom}({\bf S}^{3})=\mathrm{SO(4)}$   & $\text{Isom}({\bf E}^{3})=\mathrm{ISO(3)}$ & $\text{Isom}({\bf H}^{3})=\mathrm{SO(3,1)}$\\[6pt]  
  \hline
   \end{tabular}
} 
\end{table}


For all values of the parameters $\alpha,\chi$ the  Lie algebra \eqref{jj}  has two quadratic Casimir elements, which are given by
\begin{align}
&C_1=\alpha\,P_0^2+P_1^2+P_2^2+\chi\,(\alpha\,J_0^2+J_1^2+J_2^2),\nonumber \\
&C_2=\tfrac 12 \left( \alpha\,(J_0\,P_0+P_0\,J_0) +J_1\,P_1+P_1\,J_1+ J_2\,P_2+P_2\,J_2 \right).\label{cas}
\end{align}
This implies that  the space of  Ad-invariant symmetric bilinear forms of this Lie algebra is  two-dimensional. If one identifies  the duals of $J_a$ and $P_a$ with, respectively, $P_a$ and $J_a$, the pairings corresponding to 
 $C_1$ and $C_2$ are given, in this order,  by 
  \begin{align}
    &  \langle J_a,P_b\rangle_s=0, & 
   &\langle J_a,J_b\rangle_s= g_{ab}, &
   &\langle P_a,P_b\rangle_s= \chi\, g_{ab}.  
   \nonumber\\
 & \langle J_a,P_b\rangle_t=g_{ab}, &  &\langle J_a,J_b\rangle_t=0, &   &\langle P_a,P_b\rangle_t=0,  \label{pair}
  \end{align}
  with $g=\text{diag}(-1,1,1)$ for Lorentzian signature and $g=\text{diag}(1,1,1)$ in the Euclidean cases. 
These symmetric Ad-invariant bilinear forms  were first considered in the context of (2+1)-gravity by Witten~\cite{Witten1}. 
It is shown in~\cite{Witten1} that the second pairing is the one  appropriate for (2+1)-gravity in the sense that it allows  one
 to reformulate  (2+1)-gravity as a Chern-Simons gauge theory with the relevant isometry group as a gauge group.  The Chern-Simons gauge  field  is then given by
$
A=e^aP_a+\omega^aJ_a
$, where $e$ is the triad and $\omega$ the spin connection in Cartan's formulation of (2+1)-gravity. If one takes   $\langle\cdot, \cdot \rangle_t$ as the Ad-invariant symmetric bilinear form in the Chern-Simons action, then for all values of the cosmological constant and the signature one obtains   the Einstein-Hilbert  action for (2+1)-gravity~\cite{Witten1}.

The other quadratic Casimir element yields a gauge theory with the same equations of motion but with a different symplectic structure.
It is remarkable, see~\cite{Witten1}, that only the pairing $\langle \cdot, \cdot \rangle_s$   can be generalised to (3+1)-dimensions. Although its relation to (2+1)-gravity is subtle \cite{MSquat, MS, etera}, this provides a strong motivation to consider this pairing as well.


\sect{Drinfel'd double structures}

A $2d$-dimensional Lie algebra $\mathfrak{a}$ has the  structure of a (classical) Drinfel'd double  \cite{Drinfel'da}   if there exists a basis $\{X_1,\dots,X_d,x^1,\dots,x^d \}$ of $\mathfrak a$ in which the 
Lie bracket takes
the form
\begin{align}
[X_i,X_j]= c^k_{ij}X_k, \qquad  
[x^i,x^j]= f^{ij}_k x^k, \qquad
[x^i,X_j]= c^i_{jk}x^k- f^{ik}_j X_k  \;.\label{agd}
\end{align}
This implies that the two sets of generators $\{X_1,\dots,X_d\}$ and $\{x^1,\dots,x^d \}$ form two Lie subalgebras with structure constants  $c^k_{ij}$ and $f^{ij}_k$, respectively. Moreover,  the expression for the mixed brackets $[x^i,X_j]$ implies  that an  Ad-invariant quadratic form on $\mathfrak{a}$ is given by
\begin{align}\label{ages}
 \langle X_i,X_j\rangle= 0,\qquad \langle x^i,x^j\rangle=0, \qquad
\langle x^i,X_j\rangle= \delta^i_j,\qquad \forall i,j  ,
\end{align}
and  a quadratic Casimir operator for $\mathfrak{a}$ by
\begin{align}
C=\tfrac12\sum_i{(x^i\,X_i+X_i\,x^i)}. 
\label{cascas}
\end{align}
A Lie algebra with a Drinfel'd double (DD) structure  can therefore be viewed as a pair of Lie algebras of the same dimension with a specific set of crossed commutation rules that guarantee the existence of  the  Ad-invariant symmetric bilinear form \eqref{ages}.
In the sequel, we will refer to Lie algebras with a DD structure as DD Lie algebras.

The connection between DD Lie algebras and quantum deformations arises  from the fact that the former provide the Lie bialgebra structures of  DD quantum groups.  Each quantum universal enveloping algebra  $(U_z(\mathfrak g),\Delta_z)$ of a Lie algebra  $\mathfrak g$  is associated with  a
unique Lie bialgebra structure
$(\mathfrak g,\delta)$.   The cocommutator $\delta$ is given by the skew-symmetric part of the first-order of the coproduct  $\Delta_z$ in  the deformation parameter $z$:
\be
\delta (X)=\frac 1 2\left(\Delta_z (X)-\sigma\circ \Delta_z (X)\right) +\mathcal O(z^2),\qquad
\forall X\in \mathfrak g ,
\ee
where $ \sigma$ is the flip operator  $ \sigma (X\otimes
Y)=Y\otimes X$. 
Given a 
 Lie bialgebra ($\mathfrak{g},\delta$) and
a basis $\{X_i\}$  of $\mathfrak{g}$, one therefore obtains a DD Lie algebra as follows. The Lie bracket and the cocommutator of $\mathfrak g$ define the structure constants
($f^{lm}_n,c^k_{ij}$)  as
\be
[X_i,X_j]= c^k_{ij}X_k, \qquad  \delta(X_n)=f_{n}^{lm} X_l\otimes X_m.
\label{qqa}
\ee
Note  that the quantum deformation parameter $z$ is included in the structure constants  $f$.
The cocycle condition for the cocommutator $\delta$ then takes the form of a 
 compatibility condition between the
structure constants $c$ and $f$
\be
f^{ab}_k c^k_{ij} = f^{ak}_i c^b_{kj}+f^{kb}_i c^a_{kj}
+f^{ak}_j c^b_{ik} +f^{kb}_j c^a_{ik}  \;. \label{agb}
\ee  
Denoting by  $\{x^i\}$ the  basis of   $\mathfrak{g} ^*$ 
dual to $\{X_i\}$, that is,
\begin{align}\label{pairdd}
  \langle X_i,X_j\rangle= 0,\qquad \langle x^i,x^j\rangle=0, \qquad
\langle x^i,X_j\rangle= \delta^i_j,\qquad \forall i,j   ,
\end{align}
one can then show by a direct computation
 that ($\mathfrak{g} ^*,\delta^\ast$) is also a Lie
bialgebra with `interchanged' structure constants
\be
[x^i,x^j]=f^{ij}_k x^k, \qquad \delta^\ast(x^n)=c^{n}_{lm} x^l\otimes x^m  .
\label{agc}
\ee
This duality  leads naturally to the consideration that
the pair ($\mathfrak{g} ,\mathfrak{g}^*$), and the  associated vector space 
$\mathfrak{a}=\mathfrak{g} \oplus \mathfrak{g}^*$ can be endowed with a Lie algebra
structure by means of the brackets (\ref{agd}) and that the dual relations (\ref{pairdd}) provide the 
 an  Ad-invariant quadratic form on $\mathfrak{a}$.

Consequently, 
the resulting  Lie algebra $\mathfrak{a}\equiv D ({\mathfrak{g}})$ is called the
 DD Lie algebra of $(\mathfrak{g},\delta)$ and obviously coincides with the DD Lie algebra of $(\mathfrak{g}^\ast,\delta^\ast)$. Its Lie group  is the DD Lie group associated with $(\mathfrak{g},\delta)$. The Lie algebras
$\mathfrak{g}$ and $\mathfrak{g}^*$ are Lie subalgebras of $D ({\mathfrak{g}})$, and the 
compatibility conditions (\ref{agb}) are  the Jacobi identities
for $D ({\mathfrak{g}})$.

Moreover, if $\mathfrak{g}$ is a finite-dimensional Lie algebra,
then $D ({\mathfrak{g}})$  can be endowed with a
(quasi-triangular) Lie bialgebra structure $(D ({\mathfrak{g}}),\delta_{D})$ that is determined by the canonical  classical
$r$-matrix
\be
r=\sum_i{x^i\otimes X_i} ,
\label{rcanon}
\ee
through
\be
\delta_{D}(Y)=[ Y \otimes 1+1\otimes Y ,  r],
\quad
\forall Y\in D ({\mathfrak{g}}).
\label{rcanon2}
\ee
The cocommutator $\delta_{D}$  then  takes the form
\be
\delta_{D}(x^i)=\delta^\ast(x^i)=c^i_{jk}\,x^j\otimes x^k ,\qquad
\delta_{D}(X_i)=-\delta(X_i)=- f_i^{jk}\,X_j\otimes
X_k \;,
\label{codob}
\ee
and the Lie bialgebra 
  $(\mathfrak{g},\delta)$ and its dual ($\mathfrak{g}^\ast
,\eta$) are sub-Lie bialgebras  of the Lie bialgebra $(D(\mathfrak g), \delta_{D})$.

Note  that  the cocommutator $\delta_D$  
  only depends  on the  skew-symmetric component of the $r$-matrix (\ref{rcanon}), namely
\be  r'=\tfrac12 \sum_i{x^i\wedge X_i}  .
\label{rmat}
\ee
This is due to the fact that the symmetric component of the $r$-matrix defines a canonical quadratic Casimir element of  $D(\mathfrak g)$ in the form 
(\ref{cascas}),
which implies that the associated element of the Lie algebra $D(\mathfrak g)\otimes D(\mathfrak g)$
\be\label{omega}
\Omega=r-r'=\tfrac12\sum_i{(x^i\otimes X_i+X_i\otimes x^i)},
\ee
is  invariant under the action of $D(\mathfrak g)$
\be [ Y
\otimes 1+1\otimes Y ,\Omega]=0 ,\qquad\forall Y\in D(\mathfrak g).
\ee
Obviously, the Lie algebra $D(\mathfrak g)$ may of course have other quadratic Casimir elements in addition to $\Omega$.
 
To summarise, if a Lie algebra $\mathfrak{a}$ has a DD structure (\ref{agd}),  then this implies that  $(\mathfrak{a},\delta_{D})$ is a Lie bialgebra with canonical $r$-matrix given by (\ref{rcanon}). Therefore, there exists a quantum algebra $(U_z(\mathfrak{a}),\Delta_z)$ whose first-order  coproduct is given by $\delta_{D}$, and this quantum deformation  can be viewed as  the quantum symmetry corresponding to the given DD structure for $\mathfrak{a}$. 
In the following  sections we  explore systematically all  instances of this construction for the  Lie algebras  $so(3,1)$  and  $so(2,2)$ and, furthermore, we also determine which of the so obtained DD structures  admit a  well defined limit where the cosmological constant tends to zero.


\sect{ The Lie algebra $so(3,1)$ as a Drinfel'd double}

In this section we present a systematic investigation of all DD structures on the Lie algebra ${so}(3,1)$, which corresponds to  the isometry group
 of either  the (2+1)-dimensional
de Sitter space ${\bf dS}^{2+1}$ or the three-dimensional hyperbolic space ${\bf H}^{3}$.

For this purpose we   consider   the complete classification of  the six-dimensional  DD Lie bialgebras   given in~\cite{Snobl}  which, in turn,  corresponds to the classification of three-dimensional real Lie bialgebras obtained in~\cite{gomez}.  The results  in \cite{gomez,Snobl}
show that there exist {\em four}  three-dimensional  Lie bialgebras whose DD Lie algebra is isomorphic to ${so}(3,1)$. Note that, 
 although some of  the Lie bialgebras  in~\cite{Snobl} become isomorphic for specific values of the deformation parameter, since we are interested in $\eta$ as a continuous deformation parameter then these four cases have to be considered as essentially distinct deformations.
For each of these four DD structures, we will derive a  basis transformation from the canonical basis (\ref{agd}) for the  DD Lie bialgebra  to the kinematical basis  \eqref{jj} thus  identifying the admissible values of the parameters $\alpha$ and $\chi$, that is,   the metric and the cosmological constant. Next we will determine the Ad-invariant symmetric bilinear form (\ref{pairdd}) and analyse it in relation with the two pairings (\ref{pair}). 
Finally we will express the canonical  classical $r$-matrix (\ref{rcanon}) in the kinematical basis, and we will sketch 
the main features of the associated  quantum deformation.

We stress that the following results
are based on the {\em complete} classification of 
DD structures for the  Lie algebras $so(3,1)$ and $so(2,2)$. As the classical $r$-matrices determine the corresponding quantum deformations uniquely, this amounts to a characterisation of all  DD quantum deformations of the (Anti) de Sitter algebra that are consistent with (2+1)-gravity context.


\subsection{Case A: an $so(3,1)$-DD from a quantum deformation of $so(2,1)$}

This DD corresponds to case 2 in~\cite{gomez} and case $(8 | 5.ii | \lambda)$ in~\cite{Snobl}.  The   DD structure is induced by the Lie bialgebra associated to the  standard (or quasi-triangular) Drinfel'd-Jimbo quantum deformation of $\mathfrak{g}\equiv so(2,1)$ with real deformation parameter $\eta\equiv \lambda\neq 0$~\cite{Drinfel'da,Jimbo}. The Lie brackets for $\mathfrak{g}$ and $\mathfrak{g}^\ast$ are given by
\begin{align}
&[X_0,X_1]= X_2,
&
 &[X_0,X_2]=  -X_1,
&
&[X_1,X_2]=-\,X_0,\nonumber\\
&[x^0,x^1]= -\eta\,x^1 ,
&
&[x^0,x^2]=-\eta\,x^2,
&
&[x^1,x^2]=0,
\label{xa}
\end{align}
which implies that the crossed relations take the form
\begin{align}
&[x^0,X_0]=0,
&
&[x^0,X_1]= -x^2+\eta\,X_1 ,
&
&[x^0,X_2]=x^1+\eta\,X_2, \nonumber \\
&[x^1,X_0]=-x^2,
&
&[x^1,X_1]=-\eta\,X_0,
&
&[x^1,X_2]=x^0, \label{xb}\\
&[x^2,X_0]=x^1,
&
&[x^2,X_1]= -x^0,
&
&[x^2,X_2]=-\eta\,X_0.\nonumber
\end{align}
We will now construct  the isomorphism between the Lie algebra $\mathfrak{a}=\mathfrak{g} \oplus \mathfrak{g}^*$  and  ${so}(3,1)$ in terms of the kinematical basis  $\{J_a, P_a\}_{a=0,1,2}$ given by \eqref{jj}, which  is directly related to the isometries of the constant curvature spaces in (2+1)-gravity. 
Such a Lie algebra  isomorphism  is given by
\begin{align}
&J_0=X_0 , & 
&J_1=\tfrac{1}{\sqrt{2}}(X_1- X_2)  ,&  
&J_2=\tfrac{1}{\sqrt{2}} (X_1+ X_2)   , \label{csbasis2}\\
&P_0=-x^0 ,& 
&P_1=\tfrac{1}{\sqrt{2}} \left(\eta(X_1+X_2) +x^1- x^2\right),&  
&P_2=\tfrac{1}{\sqrt{2}} \left(\eta(X_2-X_1) +x^1+ x^2\right).
\nonumber
\end{align}
By inserting these expressions into the Lie brackets (\ref{xa}) and (\ref{xb}), one obtains the brackets in  terms of the kinematical basis
 \be 
\begin{array}{lll}
 [J_0,J_1]=\,J_2, & \quad [J_0,J_2]=-\,J_1,  & \quad  [J_1, J_2]=-\,J_0,  \\[2pt]
 [J_0,P_0]=0 ,& \quad [J_0,P_1]= \,P_2 ,& \quad [J_0, P_2]= -\,P_1,\\[2pt]
[J_1,P_0]=-\,P_2 ,& \quad [J_1,P_1]=0 ,& \quad [J_1, P_2]=-\,P_0,\\[2pt]
[J_2,P_0]=\,P_1 ,& \quad[J_2,P_1]= \,P_0 ,& \quad[J_2, P_2]=0,\\[2pt]
[P_0,P_1]=-\eta^2\,  J_2, & \quad[P_0, P_2]=-\eta^2\, J_1  ,& \quad  [P_1,P_2]= \eta^2\,  J_0   .
 \end{array}
\ee 
These expressions  correspond to a bracket of the form \eqref{jj} with $\alpha=-1$ and $\chi=-\eta^2$, that is, with
the Minkowski metric  $g=\text{diag}(-1,1,1)$  and positive cosmological constant  $\Lambda=-\chi=\eta^2>0$. 
Consequently, the Lie algebra is $\mathfrak{a}=\mathfrak{g} \oplus \mathfrak{g}^*\simeq {so}(3,1)$,  the relevant model spacetime is ${\bf dS}^{2+1}$ and 
the 
deformation parameter $\eta$ in the initial ${so}(2,1)$ Lie bialgebra provides the cosmological constant for the (2+1)-gravity model.

Furthermore,  by inserting the expressions \eqref{csbasis2} into the canonical pairing \eqref{pairdd}  from the  DD structure,
one obtains
\be
  \langle J_a,P_b\rangle=g_{ab},\qquad \langle J_a,J_b\rangle=\langle P_a,P_b\rangle=0,
  \label{xc}
\ee
which   coincides with  the pairing $\langle \cdot,\cdot\rangle_t$ in (\ref{pair}).  
  This DD  structure is therefore  appropriate for the  Chern-Simons formulation of (2+1)-gravity 
  with Lorentzian signature and positive cosmological constant.

The canonical Poisson-Lie structure on the $\mathrm{SO}(3,1)$  induced by this DD structure is determined by the  classical $r$-matrix  \eqref{rcanon}. In order to compute the corresponding expression in terms of the kinematical basis,   one needs  the inverse of the basis transformation~\eqref{csbasis2}, namely
\begin{align}
&X_0=J_0 ,& 
&X_1=\tfrac{1}{\sqrt{2}} (J_1+J_2), &  
&X_2=\tfrac{1}{\sqrt{2}} (J_2-J_1)   , \label{csbasisinv2}\\
&x^0=-P_0,& 
&x^1=\tfrac{1}{\sqrt{2}}  (P_1+P_2 + \eta(J_1-J_2)),&  
&x^2=\tfrac{1}{\sqrt{2}}  (P_2- P_1 + \eta(J_1+J_2)).
\nonumber
\end{align}
Inserting these expressions into \eqref{rcanon}, one finds
\be
r_{\rm A} =  \sum_{i=0}^2 x^i\otimes X_i=   
\eta\,J_1\wedge J_2 + (-P_0\otimes J_0 +P_1\otimes J_1 +P_2\otimes J_2).
\label{xd}
\ee
The  Casimir (\ref{cascas}) for this DD Lie algebra and the invariant element $\Omega$ (\ref{omega}) read \begin{align}
& C=\tfrac12\sum_{i=0}^2{(x^i\,X_i+X_i\,x^i)} =\tfrac 12 \left( -J_0\,P_0  -P_0\,J_0 +J_1\,P_1 +P_1\,J_1  +J_2\,P_2 +P_2\,J_2\right)\equiv C_2, \cr
&
\Omega= \tfrac12\sum_{i=0}^2{(x^i\otimes X_i+X_i\otimes x^i)}\cr
&\;\;\; =\tfrac 12 \left( -J_0\otimes P_0  -P_0\otimes J_0 +J_1\otimes P_1 +P_1 \otimes J_1  +J_2 \otimes P_2 +P_2 \otimes J_2 \right).
\label{casom}
\end{align}
Hence $\Omega$   coincides with the tensorised Casimir $C_2$  (\ref{cas}) (for $\alpha=-1$) and by subtracting this term from (\ref{xd}), one obtains the skew-symmetric $r$-matrix  $r'_{\rm A} = r_{\rm A} -\Omega$ given by
\begin{align} 
r'_{\rm A} 
=&\eta\,J_1\wedge J_2+ \tfrac{1}{2} (-P_0\wedge J_0+
P_1\wedge J_1 + P_2\wedge J_2).
\label{xe}
\end{align}
This shows that the associated  quantum deformation of $so(3,1)$  would be the one induced by the standard deformation on the (2+1)-dimensional Lorentz algebra $sl(2,\mathbb R)\simeq so(2,1)$~\cite{Drinfel'da,Jimbo} generated by $\eta\,J_1\wedge J_2$,  plus three Reshetikhin  twists $P_a\wedge J_a$. 
This turns out to be just the DD structure investigated in \cite{BHM}, where it was constructed from a `hybrid'  two-parametric Lie bialgebra  of $sl(2,\RR)\simeq so(2,1)$~\cite{hybrida,hybridb}  with the two parameters  $(\eta,z)$ corresponding to the cosmological constant $(\eta)$ and the signature of the metric $(z)$. Note, however,  that the second deformation parameter $z$ can be reabsorbed by a rescaling of the generators for $\eta\neq 0$, which yields to the $r$-matrix obtained above.

In the limit  $\eta\to 0$,  this DD Lie algebra becomes the (2+1)-dimensional Poincar\'e algebra  $D(\mathfrak g)=iso(2,1)$, which is the isometry algebra of (2+1)-dimensional  Minkowski space ${\bf M}^{2+1}$. The associated quantum deformation would be determined by the classical $r$-matrix 
 \be\lim_{\eta\to 0}r'_{\rm A} =\tfrac 1 2 (-P_0\wedge J_0+P_1\wedge J_1+P_2\wedge J_2).
 \label{xf}
 \ee
which is a sum of  three twists (see~\cite{Luktwists}).

Via the the same procedure,  the remaining DD structures on $so(3,1)$ can be expressed in terms of the  kinematical basis \eqref{jj}. In each case, the key point is  to find an appropriate  Lie algebra isomorphism from the DD basis to the kinematical one.


\subsection{Case B:   an $so(3,1)$-DD from a quantum deformation of $so(3)$}

The second DD structure for $so(3,1)$ is given by  case 4 in~\cite{gomez} and case $(9 | 5| \lambda)$ in~\cite{Snobl}. It corresponds to the Lie bialgebra underlying the standard quantum deformation of $\mathfrak{g}\equiv so(3)$ with  deformation parameter $\eta\equiv \lambda\neq 0$~\cite{Drinfel'da,Jimbo}. The Lie subalgebras $\mathfrak{g}$ and $\mathfrak{g}^\ast$ are given by
\begin{align}
&[X_0,X_1]= X_2,
&
 &[X_0,X_2]=  -X_1,
&
 &[X_1,X_2]=X_0,\cr
&[x^0,x^1]= -\eta\,x^1 ,
&
&[x^0,x^2]=-\eta\,x^2,
&
&[x^1,x^2]=0,
\label{xff}
\end{align}
and the mixed brackets read
\begin{align}
&[x^0,X_0]=0,
&
&[x^0,X_1]=  x^2+\eta\,X_1 ,
&
&[x^0,X_2]=-  x^1+\eta\,X_2,\cr
&[x^1,X_0]=-x^2,
&
&[x^1,X_1]=-\eta\,X_0,
&
&[x^1,X_2]=x^0, \label{xg}\\
&[x^2,X_0]=x^1,
&
&[x^2,X_1]= -x^0,
&
&[x^2,X_2]=-\eta\,X_0.
\nonumber
\end{align}
In this case the Lie algebra isomorphism from the DD basis to the kinematical one is 
\begin{align}
&J_0=X_0 ,& 
&J_1=\tfrac{1}{\sqrt{2}}(X_1- X_2) ,&  
&J_2=\tfrac{1}{\sqrt{2}} (X_1+ X_2)   , \label{csbasis3}\\
&P_0=x^0,& 
&P_1=\tfrac{1}{\sqrt{2}} \left(-\eta(X_1+X_2) +(x^1- x^2)\right),&  
&P_2=\tfrac{1}{\sqrt{2}}\left(\eta(X_1-X_2) +(x^1+ x^2)\right).
\nonumber
\end{align}
By introducing  these expressions into  (\ref{xff}) and (\ref{xg})  we obtain  a 
 Lie algebra   of the form \eqref{jj} with $\alpha=1$ and $\chi=-\eta^2$. This means that the associated model spacetime is three-dimensional hyperbolic space ${\bf H}^3$,
   the metric is Euclidean  $g=\text{diag}(1,1,1)$ and the cosmological constant $\Lambda=\chi=-\eta^2<0$. In the kinematical basis,  the canonical pairing \eqref{pairdd} becomes
\be
  \langle J_a,P_b\rangle=g_{ab},\qquad \langle J_a,J_b\rangle=\langle P_a,P_b\rangle=0,
  \label{xi}
\ee
which, as  in case (\ref{xc}), coincides with the pairing $\langle\cdot,\cdot\rangle_t$ in (\ref{pair}). This DD structure therefore belongs to the Chern-Simons formulation of (2+1)-gravity with Euclidean signature and negative cosmological constant.

To compute the classical $r$-matrix, one inserts the inverse  change of basis 
\begin{align}
&X_0=J_0 ,& 
&X_1=\tfrac{1}{\sqrt{2}} (J_1+J_2), &  
&X_2=\tfrac{1}{\sqrt{2}} (-J_1+J_2)   , \label{csbasisinv3}\\
&x^0=P_0,& 
&x^1=\tfrac{1}{\sqrt{2}}  \left(P_1+ P_2 + \eta(-J_1+J_2)\right),&  
&x^2=\tfrac{1}{\sqrt{2}}  \left(-P_1+ P_2 - \eta(J_1+J_2)\right),
\nonumber
\end{align}
 into the canonical classical $r$-matrix  
(\ref{rcanon}), which yields
\be
r_{\rm B} =-
\eta\,J_1\wedge J_2 + (P_0\otimes J_0 +P_1\otimes J_1 +P_2\otimes J_2).
\label{xii}
\ee

Again, the canonical Casimir element  (\ref{cascas})  for the DD corresponds  to the Casimir element  $C_2$ (\ref{cas}) and the invariant element $\Omega$   (\ref{omega}) provides its tensorised form (i.e.~the expressions (\ref{casom})  for $\alpha=1$). Subtracting the latter from the classical $r$-matrix \eqref{xii} we get
\begin{align}
r'_{\rm B} =&-\eta\,J_1\wedge J_2+ \tfrac{1}{2} (P_0\wedge J_0+
P_1\wedge J_1 + P_2\wedge J_2).
\label{xj}
\end{align}
This closely resembles the result obtained in case A (\ref{xe}) and can be regarded as its Euclidean counterpart, which was also investigated in \cite{BHM}. In the same manner, the complete quantum deformation would be provided by the standard quantum deformation generated by $-
\eta\,J_1\wedge J_2$ on the $so(3)$ Lie subalgebra of rotations, plus three Reshetikhin  twists $P_a\wedge J_a$. Both the pairing and the classical $r$-matrix exhibit a well defined limit $\eta\to 0$ ($\Lambda\to 0$) 
\be
\lim_{\eta\to 0} r'_{\rm B}=\tfrac{1}{2} (P_0\wedge J_0+
P_1\wedge J_1 + P_2\wedge J_2).
\label{xk}
\ee
which corresponds to the three-dimensional Euclidean algebra  $D(\mathfrak g)=iso(3)$, which is the isometry algebra of (2+1)-dimensional  Euclidean space  ${\bf E}^{3}$.


\subsection{Case C:  an  $so(3,1)$-DD from a quantum deformation of $i
so(2)$}

The third possibility for a DD deformation of $so(3,1)$ is given by case 9 in~\cite{gomez} and case $(7_0 | 5.ii | \lambda)$ in~\cite{Snobl}.  It is induced by the Lie bialgebra associated to a quantum deformation of the two-dimensional Euclidean algebra $\mathfrak{g}\equiv   iso(2)$, and again depends on one essential parameter $\eta\equiv \lambda\neq 0$:
\begin{align}
 &[X_0,X_1]= X_2 ,&
 &[X_0,X_2]=  -X_1,
&  &[X_1,X_2]=0,\cr
 &[x^0,x^1]= -\eta\,x^1 ,
&
&[x^0,x^2]=-\eta\,x^2,
&
&[x^1,x^2]=0. 
\label{xl}
\end{align}
The crossed relations read
\begin{align}
&[x^0,X_0]=0,
&
&[x^0,X_1]=\eta\,X_1 ,
&
&[x^0,X_2]=\eta\,X_2,\cr
&[x^1,X_0]=-x^2,
&
 &[x^1,X_1]=-\eta\,X_0,
&
&[x^1,X_2]=x^0,\label{xm}\\
&[x^2,X_0]=x^1,
&
&[x^2,X_1]= -x^0,
&
&[x^2,X_2]=-\eta\,X_0 .\nonumber
\end{align}
Note that this DD Lie algebra can be viewed as  a limiting DD structure  of cases A ({\ref{xa}) and B  (\ref{xff}) leading to the   bracket $[X_1,X_2]=0$ in (\ref{xl})  through the contraction  sequence $so(2,1)\rightarrow iso(2)\leftarrow so(3)$. However, as we shall see in the sequel, the structure of the DD kinematical algebra is {different} and cannot be obtained by `deforming' the ones in cases A and B.

To obtain a Lie algebra isomorphism from this DD Lie algebra to the kinematical one \eqref{jj}, we consider the cases $\eta<0$ and $\eta>0$ separately. For $\eta<0$, the kinematical generators are given by
 \begin{align}
&J_0=\frac{1}{\sqrt{2|\eta|}}(X_2 - x^1) ,&  
&J_1=\frac{1}{\sqrt{2|\eta|}} (X_2 + x^1)  ,&J_2=- \frac 1 {|\eta|} x^0  , \cr
&P_0=\sqrt{\frac{ |\eta|}{{2}}} (X_1 - x^2),&  
&P_1=\sqrt{\frac{ |\eta|}{{2}}} (X_1 + x^2) ,&P_2=-|\eta| X_0 .
\label{csbasis4}
\end{align}
 This corresponds to the Lie algebra  $so(3,1)$   \eqref{jj} with $\alpha=-1$ and  $\chi=-\eta^2$. Therefore, this DD structure corresponds to  ${\bf dS}^{2+1}$, the Minkowski
  metric  $g=\text{diag}(-1,1,1)$ and  positive  cosmological constant $\Lambda=\eta^2>0$, as in case A.
The canonical pairing \eqref{pairdd} becomes
\be
  \langle J_a,P_b\rangle=g_{ab},\qquad \langle J_a,J_b\rangle=\langle P_a,P_b\rangle=0,
  \label{xn}
\ee
which, once more, coincides with the pairing $\langle\cdot,\cdot\rangle_t$ in (\ref{pair}). Consequently,   this DD structure defines a second possible quantum deformation for the Chern-Simons formulation of (2+1)-gravity with Lorentzian signature and positive cosmological constant.

 The inverse change of basis of (\ref{csbasis4}) reads
\begin{align}
&X_0=- \frac 1 {|\eta|} P_2, & 
&X_1=\frac{1}{\sqrt{2|\eta|}} (P_1+P_0), &  
&X_2=\sqrt{\frac{|\eta|}{{2}}} (J_1+J_0)   , \cr
&x^0=-|\eta| J_2,& 
&x^1=\sqrt{\frac{|\eta|}{{2}}}  (J_1-J_0),&  
&x^2=\frac{1}{\sqrt{2|\eta|}}  (P_1-P_0),
\label{csbasisinv4}
\end{align}
which yields the classical   $r$-matrix 
\be
  r_{\rm C}=\tfrac{1}{2} (
J_1\wedge P_0 - J_0 \wedge P_1)+ J_2\otimes P_2 + \tfrac{1}{2} \left( -J_0\otimes P_0 - P_0\otimes J_0+
J_1\otimes P_1 + P_1\otimes J_1\right).
\label{xo}
\ee
The  Casimir (\ref{cascas})    and the invariant element $\Omega$ (\ref{omega}) for this DD Lie algebra turn out to be the same as in
(\ref{casom}),  and  we obtain  the corresponding skew-symmetric $r$-matrix
\begin{align}
r'_{\rm C} =&\tfrac{1}{2} (
J_1\wedge P_0 - J_0 \wedge P_1+J_2\wedge P_2).
\label{xp}
\end{align}

In the case $\eta>0$, the kinematical basis is given by
 \begin{align}
&J_0=\frac{1}{\sqrt{2\eta}}(X_1 - x^2) ,&  
&J_1=\frac{1}{\sqrt{2\eta}} (X_1 + x^2)  ,
&J_2= \frac 1 {\eta} x^0  , \cr
&P_0=\sqrt{\frac{ \eta}{{2}}} (X_2 - x^1),&  
&P_1=\sqrt{\frac{ \eta}{{2}}} (X_2 + x^1) ,
&P_2=\eta X_0 ,
\label{csbasis4bis}
\end{align}
and, following the same procedure, we again obtain the model spacetime   ${\bf dS}^{2+1}$ and  the same $r$-matrix (\ref{xp}).

The quantum deformation associated with this  DD is  the  standard deformation of  $so(3,1)$ \cite{BHOS3d} generated by $(J_1\wedge P_0 - J_0 \wedge P_1)$  twisted by  the Reshetikhin twist generated by  $J_2\wedge P_2$. Since neither the pairing nor the $r$-matrix depend on $\eta$, this DD structure has a well defined limit $\eta\to 0$, in which one obtains the  Poincar\'e algebra $iso(2,1)$ on ${\bf M}^{2+1}$  with the pairing (\ref{xn}) and $r$-matrix (\ref{xp}).


\subsection{Case D: a (quasi) self-dual   $so(3,1)$-DD}

The last  $so(3,1)$-DD structure is given  by case 16 in~\cite{gomez} and case $(7_\mu | 7_{1/\mu} | \lambda)$ in~\cite{Snobl}. It is associated to a quantum deformation of the one-parameter family of Lie algebras
\be
 [x^0,x^1]=\mu x^1 - x^2,
\qquad 
[x^0,x^2]=x^1 + \mu x^2,
\qquad
 [x^1,x^2]=0,
 \qquad
 \mu>0,
\label{xq}
\ee
with dual Lie algebra  $\mathfrak{g}^\ast$ 
\be 
 [X_0,X_1]= -\eta X_1/\mu + \eta X_2,
\qquad 
  [X_0,X_2]=-\eta X_1 - \eta X_2/\mu,
\qquad 
  [X_1,X_2]= 0.
\label{xr}
\ee
The dual $\mathfrak g^*$ thus depends also on the essential deformation parameter $\eta\equiv \lambda\neq 0$, and the Lie bialgebra structure becomes strictly self-dual for the isolated case $\mu=1$.
The  mixed brackets for  this DD  are given by
\begin{align}
&[x^0,X_0]=0,
&
 &[x^0,X_1]=-\mu X_1 - X_2,
&
&[x^0,X_2]=X_1 - \mu X_2,\cr
&[x^1,X_0]=-\eta x^1/\mu - \eta x^2,
&
&[x^1,X_1]=\eta x^0/\mu + \mu X_0,
&
&[x^1,X_2]=\eta x^0 - X_0, \label{xs}\\
&[x^2,X_0]=\eta x^1 - \eta x^2/\mu,
&
&[x^2,X_1]=-\eta x^0 + X_0,
&
&[x^2,X_2]=\eta x^0/\mu + \mu X_0.
\nonumber
\end{align}
To obtain the Lie algebra $so(3,1)$, we consider the kinematical  basis 
\begin{align}
&J_0=\aA(X_1+x^1)+\bB(X_2-x^2) ,& & P_0=-\eta\,\bB(X_1+x^1)+\eta\,\aA(X_2-x^2) ,\cr
&J_1=-\bB(X_1-x^1)+\aA(X_2+x^2) ,& &P_1=-\eta\,\aA(X_1-x^1)-\eta\,\bB(X_2+x^2) , \label{xxs}\\
&J_2=\frac{\mu X_0-\eta x^0/\mu}{\eta\upi} ,
&& P_2=\frac{ (X_0+\eta x^0)}{\upi} ,
\nonumber
\end{align}
where
\be
\aA=\frac{(1-\mu)}{2(1+\mu^2)}\sqrt{\frac \mu \eta} ,\qquad  \bB= \frac{(1+\mu)}{2(1+\mu^2)}\sqrt{\frac\mu\eta},\qquad \upi=\frac {(1+ \mu^2)}{\mu}.
\label{xxss}
\ee
By inserting these expressions into the Lie brackets (\ref{xq})--(\ref{xs}),  one obtains the Lie bracket  \eqref{jj} with $\alpha=1$ and $\chi= -\eta^2$. Hence   this DD Lie algebra 
coincides with $so(3,1)$, the metric is Euclidean 
 $g=\text{diag}(1,1,1)$ and the cosmological constant $\Lambda=-\eta^2<0$ negative. This shows that the relevant model spacetime is  ${\bf H}^{3}$.

  In the kinematical basis,  the canonical pairing reads
\be
\langle J_a, P_b\rangle=\frac{\mu(\mu^2-1)}{(1+\mu^2)^2}\, g_{ab},\quad \langle J_a,J_b\rangle=-\frac{2\mu^2}{\eta(1+\mu^2)^2} \,g_{ab},\quad\langle P_a,P_b\rangle=\frac{2\eta\mu^2}{(1+\mu^2)^2} \, g_{ab},
\ee
which  turns out to be  a superposition of the two bilinear forms $\langle\cdot ,\cdot \rangle_t$ and $\langle\cdot ,\cdot \rangle_s$ in (\ref{pair}):
\be
\langle\cdot ,\cdot \rangle=\frac{\mu(\mu^2-1)}{(1+\mu^2)^2}\langle\cdot ,\cdot \rangle_t-\frac{2\mu^2}{\eta(1+\mu^2)^2}\langle\cdot , \cdot \rangle_s ,
 \qquad
 \mu>0.
\label{xu}
\ee
Note that in the self-dual case $\mu=1$, the pairing becomes
\be
\langle\cdot ,\cdot \rangle_{\mu=1} =-\frac 1 {2\eta}\langle\cdot ,\cdot \rangle_s .
\label{zc}
\ee

The inverse change of basis of (\ref{xxs}) is given by
\begin{align}
&X_0=\eta J_2+P_2/\mu , & &x^0=\mu P_2/\eta-J_2 , \nonumber\\
&X_1=\upi(\eta \aA J_0-\eta \bB J_1-\bB P_0 -\aA P_1 ) ,& &x^1=\upi(\eta \aA J_0+\eta \bB J_1-\bB P_0 +\aA P_1 ),  \label{xxuu} \\
&X_2=\upi(\eta\bB J_0+\eta\aA J_1+\aA P_0  -\bB P_1 ) ,& &x^2=\upi(-\eta\bB J_0+\eta\aA J_1-\aA P_0 -\bB P_1 ), 
\nonumber
\end{align}
 and  by introducing these expressions into the canonical $r$-matrix (\ref{rcanon}) we obtain
\begin{align}
&r_{\rm D}=  J_0\wedge P_1-J_1  \wedge P_0 +\frac{(1+\mu^2 ) }{2\mu}\,P_2\wedge J_2+\frac{(\mu^2-1)}{2\eta\mu}(\eta^2 J_0\wedge J_1- P_0\wedge P_1)\cr
&\qquad  +\frac 1\eta\left(  P_0\otimes P_0+ P_1\otimes P_1+ P_2\otimes P_2 -\eta^2(    J_0\otimes J_0+J_1\otimes J_1+J_2\otimes J_2) \right)\cr
&\qquad  + \frac{(\mu^2-1)}{2\mu} (J_0\otimes P_0+P_0\otimes J_0+J_1\otimes P_1+P_1\otimes J_1+J_2\otimes P_2+P_2\otimes J_2).
\label{xv}
\end{align}
Now the quadratic Casimir  (\ref{cascas})  is  a linear superposition of $C_1$ and $C_2$  in (\ref{cas}) for  $\alpha=1$ and $\chi=-\eta^2$:
\be
C=\frac 1\eta \,  C_1 +  \frac{(\mu^2-1)}{ \mu}\,C_2 .
\label{xxv}
\ee
  Subtracting the corresponding tensorised  element  $\Omega$ (\ref{omega})  from  expression (\ref{xv})  gives rise to the skew-symmetric classical $r$-matrix 
\be
r'_{\rm D}=   J_0\wedge P_1-J_1  \wedge P_0  +\frac{(1+\mu^2 ) }{2\mu}\,P_2\wedge J_2+\frac{(\mu^2-1)}{2\eta\mu}(\eta^2 J_0\wedge J_1-P_0\wedge P_1).
\label{za}
\ee
This is a quite involved classical $r$-matrix whose associated quantum deformation would be the superposition of: a) the standard quantum deformation of  $so(3,1)$ \cite{BHOS3d} generated by $(J_0 \wedge P_1-J_1\wedge P_0)$, b) a    Reshetikhin twist  generated by $P_2\wedge J_2$, c) another deformation generated by   $(\eta^2 J_0\wedge J_1-P_0\wedge P_1)$.  In the self-dual case,  where $\mu= 1$, the $r$-matrix  reduces to
\be
 r'_{\rm D}= J_0\wedge P_1-J_1  \wedge P_0  + P_2\wedge J_2,
\label{zb}
\ee 
which turns out to be  the Euclidean counterpart of case C (\ref{xp}). However, its pairing \eqref{zc}
does not coincide with the Euclidean counterpart of the pairing in the case C (\ref{xn}). 

To investigate the cosmological limit of these DD structures, recall that the cosmological constant is given by  $\Lambda=\ -\eta^2$.
 For $\mu\neq 1$ the classical $r$-matrix  (\ref{za}) diverges in the limit $\eta\to 0$,  while it becomes independent of $\eta$ for $\mu= 1$.  The latter would be associated to a self-dual quantum deformation of  the Euclidean algebra $iso(3)$ on ${\bf E}^3$ with classical $r$-matrix (\ref{zb}). Nevertheless, 
    the pairing  (\ref{xu}) diverges in the limit $\eta\to 0$ for all possible values of $\mu> 0$.


\sect{The Lie algebra $so(2,2)$ as a Drinfel'd double}

In this section, we investigate the DD structures for the anti-de Sitter Lie algebra $so(2,2)$. There are  three Lie bialgebras whose DD is isomorphic to $so(2,2)$. These three cases  will be  analysed separately, following the same steps as in the previous section.


\subsection{Case E: an  $so(2,2)$-DD  from a   quantum deformation of $sl(2,\mathbb{R})$}

The first DD  structure for $so(2,2)$ corresponds to case 1 in~\cite{gomez} and case $(8 | 5.i | \lambda)$ in~\cite{Snobl}. It is the Lie bialgebra for  the standard quantum deformation of $\mathfrak{g}\equiv sl(2,\mathbb{R}) \simeq so(2,1)$ with deformation  parameter $\eta\equiv \lambda\neq 0$~\cite{Drinfel'da,Jimbo}. The Lie algebra $\mathfrak g$ and its dual $\mathfrak g^*$ are given by
\begin{align}
&[X_0,X_1]= 2\,X_1 ,
&
&[X_0,X_2]=  -2\,X_2 ,
&
&[X_1,X_2]= X_0, \nonumber\\
&[x^0,x^1]= -\tfrac 1 2 \eta\,x^1 ,
&
&[x^0,x^2]=-\tfrac 1 2 \eta\,x^2,
&
&[x^1,x^2]=0,
\label{zd}
\end{align}
and the crossed relations take the form
\begin{align}
&[x^0,X_0]=0,
&
&[x^0,X_1]=x^2+\tfrac 1 2 \eta\,X_1 ,
& 
&[x^0,X_2]=-x^1+\tfrac 1 2 \eta\,X_2, \nonumber\\
&[x^1,X_0]=2 x^1,
&
&[x^1,X_1]=-2 x^0-\tfrac 1 2 \eta\,X_0,
&
&[x^1,X_2]=0, \label{ze} \\
&[x^2,X_0]=- 2 x^2,
&
&[x^2,X_1]= 0,
&
&[x^2,X_2]=2 x^0-\tfrac 1 2 \eta\,X_0.\nonumber
\end{align}
In this case, the basis transformation that determines the kinematical generators is given by
\begin{align}
&J_0=-\tfrac12 (X_1 -X_2), & 
&J_1=\tfrac12 X_0 ,&  
&J_2=\tfrac12 (X_1 +X_2)   , \label{csbasis6}\\
&P_0=-\tfrac 1 2 \eta  (X_1 +X_2) +(x^1 - x^2),& 
&P_1=2 x^0,&  
&P_2=\tfrac 1 2 \eta  (X_1 -X_2) +(x^1 + x^2).
\nonumber
\end{align}
Inserting these expression into the Lie bracket, one obtains the Lie algebra   ${so}(2,2)$ in the form \eqref{jj} with $\alpha=-1$ and $\chi=\eta^2$. This means that that the metric is the Minkowski metric
 $g=\text{diag}(-1,1,1)$ and  the cosmological constant is negative $\Lambda=-\eta^2<0$.  
Inserting these expressions into formula \eqref{pairdd} for the canonical pairing, one obtains
\be
  \langle J_a,P_b\rangle=g_{ab},\qquad \langle J_a,J_b\rangle=\langle P_a,P_b\rangle=0,
  \label{zf}
\ee
which coincides with the pairing $\langle\;,\;\rangle_t$ in \eqref{pair}. This DD is therefore suitable for the Chern-Simons formulation of (2+1)-gravity  with negative cosmological constant and Lorentzian signature, which corresponds to ${\mathbf AdS^{2+1}}$.
The inverse change of basis  is given by (see  expression (5.11) in~\cite{MSquat}):
\begin{align}
&X_0=2 J_1 ,& 
&X_1=-J_0+J_2 ,&  
&X_2=J_0+J_2   , \label{csbasisinv6}\\
&x^0=\tfrac12 P_1,& 
&x^1=\tfrac12  (P_0 +P_2) +\tfrac 1 2 \eta(J_0 +J_2),&  
&x^2=\tfrac12  (-P_0 +P_2) +\tfrac 1 2 \eta(J_0 -J_2),
\nonumber
\end{align}
and the canonical $r$-matrix   reads
\be
r_{\rm E}=\eta \,
J_0\wedge J_2
+(-P_0\otimes J_0 + P_1\otimes J_1 + P_2\otimes J_2).
  \label{zg}
\ee
The  canonical Casimir (\ref{cascas})   and the invariant element $\Omega$ (\ref{omega}) are exactly given by (\ref{casom}). 
 Subtracting the component $\Omega$,  we obtain the purely skew symmetric $r$-matrix
\be
r'_{\rm E}=\eta \,
J_0\wedge J_2+ \tfrac{1}{2} (-P_0\wedge J_0+
P_1\wedge J_1 + P_2\wedge J_2).
\label{zh}
\ee
This means that the associated quantum deformation of  ${so}(2,2)$ would be a superposition of the standard quantum deformation of  $sl(2,\mathbb{R})$~\cite{Drinfel'da,Jimbo},  generated by $\eta \,
J_0\wedge J_2$, together with three Reshetikhin twists generated by $P_a\wedge J_a$.  This DD structure can thus be viewed as the anti-de Sitter counterpart of the DD structures for $so(3,1)$ given in cases A (\ref{xe})  and B (\ref{xj}), which were  considered in \cite{BHM} within the framework of a two-parametric deformation. 

In the limit $\eta\to 0$, the Lie algebra becomes the (2+1)-dimensional Poincar\'e algebra $iso(2,1)$, and  only the twists survive within the $r$-matrix:
\be
\lim_{\eta\to 0} r'_{\rm E}=\tfrac{1}{2} (-P_0\wedge J_0+
P_1\wedge J_1 + P_2\wedge J_2).
\label{zi}
\ee
This deformation therefore has a well defined $\eta\to 0$ limit, which coincides with that of the $r$-matrix in case A (\ref{xf}). 
In other words, the Poincar\'e deformation on ${\bf M}^{2+1}$  from (\ref{zi}) is the {\em common} limit of the  $so(2,2)$ DD  for ${\bf AdS}^{2+1}$ in (\ref{zh})  and of the $so(3,1)$ DD for ${\bf dS}^{2+1}$   in (\ref{xe}).
Moreover, the Euclidean counterpart of  (\ref{zi})  was obtained  in  case B through the limit   $\eta\to 0$, which yields the    $r$-matrix   (\ref{xk}).


\subsection{Case F: an $so(2,2)$-DD from a quantum deformation of $iso(1,1)$}

This DD corresponds to case $(6_0 | 5.iii | \lambda)$ in~\cite{Snobl} and case (11) in \cite{gomez}. (Note that there is a misprint in Table III of~\cite{gomez}, where it is stated that the resulting DD Lie algebra  is isomorphic to $sl(2,\RR)\oplus \RR^3$).  This is the Lie bialgebra structure  induced by a quantum deformation of the (1+1)-dimensional Poincar\'e algebra $\mathfrak{g}\equiv   iso(1,1)$ and depends on an essential deformation parameter $\eta\equiv\lambda\neq 0$. The Lie algebra $\mathfrak g$ and its dual $\mathfrak g^*$ are given by
\begin{align}
&[X_0,X_1]= - X_2,
&
&[X_0,X_2]=  -X_1,
& 
&[X_1,X_2]=0, \nonumber \\
&[x^0,x^1]= \eta\,x^1 ,
&
&[x^0,x^2]=\eta\,x^2,
&
&[x^1,x^2]=0,
\label{zj}
\end{align}
and the crossed relations read 
\begin{align}
&[x^0,X_0]=0,
&
&[x^0,X_1]=-\eta\,X_1 ,
&
&[x^0,X_2]=-\eta\,X_2, \nonumber\\
&[x^1,X_0]=-x^2,
&
&[x^1,X_1]=\eta\,X_0,
&
&[x^1,X_2]=x^0, \label{zk} \\
&[x^2,X_0]=-x^1,
&
&[x^2,X_1]= x^0,
&
&[x^2,X_2]=\eta\,X_0.
\nonumber
\end{align}

Similarly  to case C, we consider the cases $\eta<0$ and $\eta>0$ separately in order 
to obtain the appropriate  kinematical basis.
For   $\eta>0$ we   define  the change of basis
\begin{align}
&J_0=\frac{1}{\sqrt{2\eta}}(X_2 - x^1) ,&  
&J_1=\frac{1}{\sqrt{2\eta}} (X_2 + x^1),  &J_2=- \frac 1 \eta x^0  ,\nonumber\\
&P_0=\sqrt{ \frac{\eta}{{2}}} (X_1 - x^2),&  
&P_1=\sqrt{\frac{\eta}{{2}}} (X_1 + x^2), &P_2=-\eta X_0.  \label{csbasis7}
\end{align}
By inserting these expressions into the above Lie brackets, we obtain a Lie algebra of the form \eqref{jj} with $g=\text{diag}(-1,1,1)$ and $\chi=\eta^2$. Hence  we have the Lie algebra $so(2,2)$ with cosmological constant $\Lambda=-\eta^2$. The canonical pairing takes the form
\be
  \langle J_a,P_b\rangle=g_{ab},\qquad \langle J_a,J_b\rangle=\langle P_a,P_b\rangle=0,
\label{zzjj}
\ee
which again agrees with the pairing $\langle\;,\;\rangle_t$ in \eqref{pair}. This DD structure thus provides a  second possible quantum deformation for the  Chern-Simons formulation of (2+1)-gravity with Lorentzian signature and negative cosmological constant.

The inverse change of basis of (\ref{csbasis7}) is given by
\begin{align}
&X_0=- \frac 1 \eta P_2 ,& 
&X_1=\frac{1}{\sqrt{2\eta}} (P_1+P_0), &  
&X_2=\sqrt{\frac \eta 2} (J_1+J_0)   ,\nonumber \\
&x^0=-\eta J_0,& 
&x^1=\sqrt{\frac \eta 2}  (J_1-J_0),&  
&x^2=\frac{1}{\sqrt{2\eta}}  (P_1-P_0),
 \label{csbasisinv7}
\end{align}
and allows one to compute the   canonical $r$-matrix, namely
\be
  r_{\rm F}=\tfrac{1}{2} (
J_1\wedge P_0 - J_0  \wedge P_1 )
+J_2\otimes P_2+ \tfrac{1}{2} \left( -J_0\otimes P_0 - P_0\otimes J_0+
J_1\otimes P_1 + P_1\otimes J_1\right).
 \label{zl}
\ee
Again, the  Casimir (\ref{cascas})    and the invariant element $\Omega$ (\ref{omega})  are given by
(\ref{casom}).   Subtracting $\Omega$  from $r_{\rm F}$, we obtain the skew-symmetric $r$-matrix
\be
r'_{\rm F} 
=\tfrac{1}{2} (
J_1\wedge P_0 - J_0 \wedge P_1 + J_2\wedge P_2) .
\label{zm}
\ee
For $\eta<0$ the change of basis 
\begin{align}
&J_0=\frac{1}{\sqrt{2|\eta|}}(X_2 + x^1) ,&  
&J_1=\frac{1}{\sqrt{2|\eta|}} (X_2 - x^1),  &J_2= \frac {1}{ |\eta|} x^0  ,\nonumber\\
&P_0=\sqrt{ \frac{|\eta|}{{2}}} (-X_1 - x^2),&  
&P_1=\sqrt{\frac{|\eta|}{{2}}} (-X_1 + x^2), &P_2=|\eta| X_0, \label{csbasis7bis}
\end{align}
yields again the lie algebra $so(2,2)$ with the same pairing (\ref{zzjj})
 and  skew-symmetric  $r$-matrix (\ref{zm}).

The resulting quantum deformation generated by the classical $r$-matrix (\ref{zm})   is a superposition of the   standard deformation of  $so(2,2)$ \cite{BHOS3d} generated by $(J_1\wedge P_0 - J_0 \wedge P_1)$ and  a Reshetikhin twist  generated by $J_2\wedge P_2$.  
We stress that the $r$-matrix (\ref{zm}) does  coincide with the one obtained in case C (\ref{xp}) for $so(3,1)$ on ${\bf dS}^{2+1}$. As 
the $r$-matrix  does not depend on $\eta$ this, in turn, means that $r'_{\rm F} =r'_{\rm C} $ is the {\em common} classical $r$-matrix for the {\em three} Lorentzian DD structures $so(2,2)$, $so(3,1)$ and $iso(2,1)$ on ${\bf AdS}^{2+1}$, ${\bf dS}^{2+1}$  and ${\bf M}^{2+1}$, which are endowed with the same pairing  \eqref{pair}.


\subsection{Case G: a (quasi) self-dual   $so(2,2)$-DD structure}

The last DD structure for $so(2,2)$ is given by case 7 in~\cite{gomez} and case $(6_a | 6_{1/a}.i | \lambda)$ in~\cite{Snobl} where $a=(\rho+1)/(\rho-1)$ and $-1<\rho<1$.  It depends on two essential parameters $\eta\equiv\lambda\neq 0$ and $\rho$,
and it can be viewed as the Lie bialgebra for the quantum deformation of the family of Lie algebras $\mathfrak{g}\equiv \tau_3(\rho)$. The Lie bracket of $\mathfrak{g}\equiv \tau_3(\rho)$ reads
\be
 [x^0,x^1]= x^1 ,
\qquad 
[x^0,x^2]=\rho\,x^2,
\qquad
 [x^1,x^2]=0 .
 \label{zn}
\ee
Its cocommutator  determines the Lie algebra structure on the dual $\mathfrak{g}^\ast$, which is given by
\be 
 [X_0,X_1]=  \eta\,X_1 ,
\qquad 
  [X_0,X_2]=  -\eta\,\rho\,X_2 ,
\qquad 
  [X_1,X_2]= 0,
  \label{zo}
\ee
and  is isomorphic to the  Lie algebra  $\mathfrak{g}^\ast\equiv  \tau_3(-\rho)$. This implies that the Lie bialgebra $\mathfrak g$  becomes strictly self-dual  when $\rho= 0$. In the limit $\rho\to 1$,  $\mathfrak{g}$ is isomorphic to the  `book' algebra $\tau_3(1)$, while the dual 
$\mathfrak{g}^\ast$   is isomorphic to the (1+1)-dimensional Poincar\'e algebra $iso(1,1)$, thus leading to  the previous case F (\ref{zj}).
  Conversely, if $\rho\to-1$ then  $\mathfrak{g}\equiv iso(1,1)$ and 
 $\mathfrak{g}^\ast\equiv \tau_3(1)$.  
  The crossed relations are given by
\begin{align}
&[x^0,X_0]=0,
&
&[x^0,X_1]=- X_1 ,
&
&[x^0,X_2]=-  \rho\,X_2,\nonumber\\
&[x^1,X_0]= \eta\,x^1,
&
&[x^1,X_1]=- \eta\,x^0 +  X_0,
&
&[x^1,X_2]=0, \label{zp}\\
&[x^2,X_0]=- \eta\rho\,x^2,
&
&[x^2,X_1]= 0,
&
&[x^2,X_2]=\rho(\eta\,x^0 +  X_0).
\nonumber
\end{align}

This DD algebra turns out to be isomorphic to $so(2,2)$ when $\rho\neq 0$ and to the direct sum $sl(2,\RR)\oplus \RR^3$ in the self-dual case $\rho=0$.   Hence hereafter we impose $\rho\neq 0$.
 A change of basis that  transforms this double into a kinematical realisation of $so(2,2)$
 is the following
\begin{align}
&J_0=\frac{1}{2\sqrt{\eta}\rho} \left(X_2 - x^2 -(X_1 + x^1)\rho\right), & & P_0=\frac{\sqrt{\eta}}{2\rho} \left(X_2 - x^2 +(X_1 + x^1)\rho\right),
\nonumber\\
&J_1=\frac{1}{2\sqrt{\eta}\rho} \left(X_2 + x^2 -(X_1 - x^1)\rho\right),& &P_1=\frac{\sqrt \eta}{2\rho} \left(X_2 + x^2 +(X_1 - x^1)\rho\right),\label{zq}\\
& J_2=\frac{1}{2{\eta}\rho} \left(X_0(\rho-1) - x^0 \eta (\rho+1)\right) ,& &P_2=\frac{1}{2\rho} \left(-X_0(\rho+1) - x^0 \eta (1- \rho)\right) .\nonumber
\end{align}
The resulting Lie bracket is of the form \eqref{jj}  with $g=\text{diag}(-1,1,1)$, $\chi=\eta^2$ and $\Lambda=-\eta^2$. Note that the parameter $\rho$ does not enter the Lie bracket and therefore affects  neither the metric nor the cosmological constant.

 The
    pairing  reads
\be
  \langle J_a,P_b\rangle= g_{ab} \frac{(1+\rho^2)}{2\rho^2},
  \qquad 
  \langle J_a,J_b\rangle=g_{ab} \frac{(1-\rho^2)}{2\eta \rho^2},
  \qquad 
  \langle P_a,P_b\rangle=g_{ab} \frac{\eta(1-\rho^2)}{2\rho^2},
  \label{zr}
\ee
which is a superposition of the two  pairings $\langle\cdot ,\cdot \rangle_t, \langle\cdot ,\cdot\rangle_s$ in \eqref{pair} whose coefficients depend on both parameters $\rho$ and $\eta$:
\be
 \langle \cdot,\cdot\rangle= \frac{(1+\rho^2)}{2\rho^2}\langle \cdot,\cdot\rangle_t 
 +\frac{(1-\rho^2)}{2\eta\rho^2}  \langle \cdot,\cdot\rangle_s
, \qquad -1<\rho<1,\, \rho\neq 0.
  \label{zs}
\ee

To compute the canonical classical $r$-matrix, we invert the change of basis
\begin{align}
&X_0=\frac 1 {2} \left(\eta J_2(1-\rho) -  P_2(1+\rho) \right) ,& &x^0=\frac{1}{2\eta} \left(-\eta(1+\rho)J_2 +  P_2(1- \rho) \right) , \nonumber\\
&X_1=\frac{1}{2\sqrt{\eta}} \left(-\eta(J_0 +J_1)+ P_0+P_1 \right) ,& &x^1=\frac{1}{2\sqrt{\eta}} \left(\eta(J_1-J_0)+ P_0-P_1 \right),  \label{ztt}\\ 
&X_2=\frac{\rho}{2\sqrt\eta} \left (\eta(J_0 +J_1)+ P_0+P_1 \right)  ,& &x^2=\frac{\rho}{2\sqrt{\eta}}  \left(\eta(J_1-J_0)- P_0+P_1 \right).
\nonumber
\end{align}
Inserting these expressions into the  $r$-matrix (\ref{rcanon}) yields
\begin{align}
&  r_{\rm G}=\frac { (1+\rho^2)} 4 (J_1 \wedge P_0 - J_0  \wedge P_1 )+\frac  \rho 2 J_2\wedge P_2 +
\frac{(1-\rho^2)}{4\eta} (\eta^2 J_0\wedge J_1 + P_0 \wedge P_1) \nonumber\\
&
\qquad -\frac{(1-\rho^2)}{4\eta}\left(-P_0\otimes P_0+ P_1\otimes P_1+ P_2\otimes P_2 + \eta^2(-J_0\otimes J_0 + J_1\otimes J_1+ J_2\otimes J_2)
\right)\nonumber\\
&\qquad
-\frac{ (1+\rho^2)}{4}(- J_0\otimes P_0 - P_0\otimes J_0 +J_1\otimes P_1 + P_1\otimes J_1+
J_2\otimes P_2 + P_2\otimes J_2) .
\label{zt}
\end{align}
By taking into account that  the quadratic Casimir  (\ref{cascas})  is the linear superposition of $C_1$ and $C_2$ (\ref{cas}) (with  $\alpha=-1$ and $\chi=\eta^2$) given by
\be
C= -\frac{(1-\rho^2)}{4\eta}\, C_1  -\frac{ (1+\rho^2)}{2}\, C_2 ,
\label{zu}
\ee
one can subtract the corresponding tensorised element $\Omega$ (\ref{omega}) and obtains  the  skew-symmetric $r$-matrix
\be
r'_{\rm G}= \frac { (1+\rho^2)} 4 (J_1 \wedge P_0 - J_0  \wedge P_1 )+\frac  \rho 2   J_2\wedge P_2 +
\frac{(1-\rho^2)}{4\eta} (\eta^2 J_0\wedge J_1 + P_0 \wedge P_1).
\label{zv}
\ee
This $r$-matrix resembles case D (\ref{za}) for $so(3,1)$ as    it
would give rise to a superposition of  the standard deformation of  $so(2,2)$ \cite{BHOS3d} generated by $(J_1 \wedge P_0-J_0\wedge P_1 )$, 
together with a  Reshetikhin twist  generated by $J_2\wedge P_2$  plus the additional deformation generated by  the term $(\eta^2 J_0\wedge J_1+P_0\wedge P_1)$. Since $-1<\rho<1$ and $\rho\neq 0$, both the classical $r$-matrix  and the pairing are divergent in the limit $\eta\to 0$.  Note also that  the limit $\rho\to + 1$ yields   the classical $r$-matrix (\ref{zm}) and the pairing  (\ref{zzjj})  from case F.


\sect{Discussion and open problems}

For all seven DD structures, the Lie algebras, pairings, skew-symmetric $r$-matrices (including their $\Lambda\to 0$ limit, in case it exists) and associated model spacetimes are presented 
 in Table \ref{table2}.
  This table  shows that $so(3,1)$ and $so(2,2)$ DD structures (and the associated quantum deformations suitable for (2+1)-gravity) can be grouped into the following three classes:

\begin{table}[t] {\footnotesize
 \noindent
\caption{{\small The seven Drinfel'd double   Lie algebras $so(3,1)$ and $so(2,2)$.    In each case the  result  corresponding to  the    limit $\eta\to 0$ ($\Lambda=0$), leading to either $iso(2,1)$ or $iso(3)$, is also indicated. In case D, for which $\mu> 0$, this limit exists only for $\mu= 1$.
In case G, recall that $-1<\rho<1,\, \rho\neq 0$.
  }}
\label{table2}

\medskip
\noindent\hfill

\begin{tabular} {lllllll}
 \hline
 & & & & & &\\[-1.5ex]
 \# & Metric & $\Lambda$ & Pairing &   Skew-symmetric $r$-matrix & $D(\mathfrak g)$& Space\\[+1.5ex]
 \hline
 & & & & & \\[-1.5ex]
 A &$ (-1,1,1)$ & $\eta^2$ & $\langle\;,\;\rangle_t$ & $r'_{\rm A}=\eta J_1\wedge J_2+\tfrac 1 2 (-P_0\wedge J_0+P_1\wedge J_1+P_2\wedge J_2)$ & $so(3,1)$&${\bf dS}^{2+1}$\\[+1.5ex]
     &  & 0 & $\langle\;,\;\rangle_t$ & $ r'_{\rm A}= \tfrac 1 2 (-P_0\wedge J_0+P_1\wedge J_1+P_2\wedge J_2)$ & $iso(2,1)$&${\bf M}^{2+1}$\\[+2.5ex]
 B & $(1,1,1) $& $-\eta^2$ & $\langle\;,\;\rangle_t$ & $r'_{\rm B}=-\eta J_1\wedge J_2+\tfrac 1 2 (P_0\wedge J_0+P_1\wedge J_1+P_2\wedge J_2)$ & $so(3,1)$&${\bf H}^{3}$ \\[+1.5ex]
   &  & 0 & $\langle\;,\;\rangle_t$ & $ r'_{\rm B}= \tfrac 1 2 (P_0\wedge J_0+P_1\wedge J_1+P_2\wedge J_2)$ & $iso(3)$&${\bf E}^{3}$\\[+2.5ex]
 C & $(-1,1,1)$ & $\eta^2$ & $\langle\;,\;\rangle_t$ &  $r'_{\rm C}=\tfrac 1 2 (J_1\wedge P_0-  J_0 \wedge P_1+J_2\wedge P_2)$   & $so(3,1)$ &${\bf dS}^{2+1}$\\[+1.5ex]
   &   & $0$ & $\langle\;,\;\rangle_t$ &  $r'_{\rm C}=\tfrac 1 2 (J_1\wedge P_0-  J_0 \wedge P_1+J_2\wedge P_2)$   & $iso(2,1)$ &${\bf M}^{2+1}$\\[+2.5ex]
 D & $(1,1,1)$ & $-\eta^2$ &  
$ \frac{\mu(\mu^2-1)}{(1+\mu^2)^2}\langle\;,\;\rangle_t$ 
 & $r'_{\rm D}=  J_0\wedge P_1-J_1  \wedge P_0  +\frac{(1+\mu^2 ) }{2\mu}\,P_2\wedge J_2   $ 
& $so(3,1)$&${\bf H}^{3}$\\[+1.5ex]
 & & & $\frac{-2\mu^2}{\eta(1+\mu^2)^2}\langle\;, \;\rangle_s$ & $\qquad\quad+\frac{(\mu^2-1)}{2\eta\mu}(\eta^2 J_0\wedge J_1-P_0\wedge P_1) $
 \\[+1.5ex]
    &   &   $0$  &   None
 & $r'_{\rm D}=  J_0\wedge P_1-J_1  \wedge P_0  + \,P_2\wedge J_2   $ \quad $(\mu= 1)$
& $iso(3)$&${\bf E}^{3}$\\[+2.5ex]
 E &$ (-1,1,1)$ & $-\eta^2$ & $\langle\;,\;\rangle_t$ & $r'_{\rm E}=\eta J_0\wedge J_2+\tfrac 1 2 (-P_0\wedge J_0+P_1\wedge J_1+P_2\wedge J_2)$ & $so(2,2)$&${\bf AdS}^{2+1}$  \\[+1.5ex]
 &  & $0$ & $\langle\;,\;\rangle_t$ & $r'_{\rm E} = \tfrac 1 2 (-P_0\wedge J_0+P_1\wedge J_1+P_2\wedge J_2)$ & $iso(2,1)$&${\bf M}^{2+1}$  \\[+2.5ex]
 F & $(-1,1,1)$ &  $-\eta^2$ & $\langle\;,\;\rangle_t$ & $r'_{\rm F}  =\tfrac 1 2 (J_1\wedge P_0- J_0 \wedge P_1 +J_2\wedge P_2) $   & $so(2,2)$&${\bf AdS}^{2+1}$ \\[+1.5ex]
    &  &  $0$ & $\langle\;,\;\rangle_t$ & $r'_{\rm F}  =\tfrac 1 2 (J_1\wedge P_0- J_0 \wedge P_1 +J_2\wedge P_2) $   & $iso(2,1)$&${\bf M}^{2+1}$ \\[+2.5ex]
  G & $(-1,1,1)$ & $-\eta^2$ & $\frac{(1+\rho^2)}{2\rho^2}\langle \cdot,\cdot\rangle_t$ & $r'_{\rm G}=  \frac { (1+\rho^2)} 4 (J_1 \wedge P_0 - J_0  \wedge P_1 )+\frac  \rho 2    J_2\wedge P_2 
$ & $so(2,2)$&${\bf AdS}^{2+1}$\\[+1.5ex]
    &   &   &$ \frac{+(1-\rho^2)}{2\eta\rho^2}  \langle \cdot,\cdot\rangle_s$   & $\qquad\quad +
\frac{(1-\rho^2)}{4\eta} (\eta^2 J_0\wedge J_1 + P_0 \wedge P_1)$&  & \\[+1.5ex]
    &  &  $0$ & None & None   & $iso(2,1)$&${\bf M}^{2+1}$ \\[+1.5ex]
\hline
 \end{tabular}
\hfill}
\end{table}

\begin{itemize}

\item 1) {\bf Cases A-B and E}. The DD structure  A  corresponds to Lorentzian (2+1)-gravity with Lorentzian signature and positive cosmological constant, while case B corresponds to its Euclidean analogue with negative cosmological constant. Case E is just the anti-de Sitter counterpart of case A. These three  cases admit a well defined $\Lambda\to 0$ limit, in which they give rise  to the pairing $\langle\cdot,\cdot\rangle_t$ and to either a  DD structure for the Poincar\'e algebra $iso(2,1)$ (cases A and E) or to a  DD structure  on the Euclidean algebra $iso(3)$ (case B). These three DD structures were investigated in \cite{BHM}.
 
\item 2) {\bf Cases C and F}. The  pairing, $r$-matrix and DD structure of case F are the anti-de Sitter  counterparts of the ones in case C. This is a new class of DD
structures for Lorentzian  (2+1)-gravity which is valid for all values of the cosmological constant and admits a well defined $\Lambda\to 0$, which yields $iso(2,1)$. However, in contrast to the DD structure in cases A and E, which have case B as their Euclidean counterpart, the extension 
  of the DD structure from cases C and F to Euclidean (2+1)-gravity with negative cosmological constant does not seem possible.  
  This problem could possibly be overcome via  analytic continuation~\cite{BHM} or graded contraction techniques~\cite{BHOS3d,LBC}. 
  
  It is worth emphasising that cases  C and F are   related to   $\kappa$-deformations of the (anti-)de Sitter algebras constructed in~\cite{BHOS3d} (see also  \cite{amel, LBC,Tallin}). In fact, the first part $(J_1\wedge P_0-  J_0 \wedge P_1)$ of the $r$-matrix generates the $\kappa$-deformation, but in order to obtain a deformation consistent 
  with Chern-Simons formulation of (2+1)-gravity, the additional twist term generated by $J_2 \wedge P_2$ has to be added. This statement remains true  in the limit  $\Lambda\to 0$.

\item 3) {\bf Cases D and G}. Both of them present a `mixed' pairing and a similar form for the $r$-matrices, although the coefficients involved in the mixing are different. Therefore, it is not possible to interpret case D as the Euclidean  counterpart   of case G. Nevertheless, the two cases exhibit strong similarities  insofar as they are the only cases endowed with a pairing that is a superposition of the two pairings $\langle\cdot,\cdot\rangle_t$ and $\langle\cdot,\cdot\rangle_s$, and they do not possess  a well defined $\Lambda\to 0$ limit. As their pairings do in general not coincide with the pairing  $\langle\cdot,\cdot\rangle_t$ for the Chern-Simons formulation of (2+1)-gravity, the interpretation of these DD structures in the context of (2+1)-gravity is subtle. 

\end{itemize}


To summarise, the results in this article show that there are essentially two possible types of DD structures on the de Sitter algebra $so(3,1)$ and the anti-de Sitter algebra $so(2,2)$ that admit well defined cosmological limits and hence define a DD structure on the (2+1)-dimensional Poincar\'e algebra $iso(2,1)$.  While it was shown in~\cite{MSquat,MS} that simple $\kappa$-deformations are not compatible with the Chern-Simons formulation of (2+1)-gravity, the 
 DD structures from cases A, B, C, E and F in this paper  exhibit the appropriate  pairing and hence are suitable candidates for the kinematical quantum group symmetries of (2+1)-gravity.

The deformations  in cases C and F are related to $\kappa$-Poincar\'e symmetries and their generalisations to non-vanishing cosmological constant. However,  in all cases the classical $r$-matrices involve {\em additional} terms that are not present in the $\kappa$-deformations. Even if these additional terms are twists,  they can be expected to have an impact in the associated non-commutative spacetimes, to  modify the particle interactions in multi-particle models and the effective symmetries of models with boundaries. 

This is due to the fact that they  do not affect the commutation rules of the associated quantum group but they do modify its coproduct. As the particle interaction in multi-particle models is governed by the coproduct and the associated non-commutative spaces are based on the dual quantum group,  the  presence of twists will  affect these constructions. Their effect  is worth to be investigated more deeply. For the case of $\kappa$-Poincar\'e models, the impact of twists has been studied in detail in \cite{Dasz}.

The  classical  DD structures  constructed  in these paper allow one to implement the cosmological constant $\Lambda$ as a deformation parameter and to realise the  limit $\Lambda\to 0$ as a  Lie bialgebra contraction~\cite{LBC}. 
This  constitutes a first step towards the construction of multi-parametric models for quantum gravity in which the cosmological limit, the semiclassical limit and the low velocity limit can be investigated separately. 

A complete analysis would require the construction of the full Hopf algebra for each DD Lie algebra in the kinematical basis,   of the associated non-commutative spacetimes as well as a careful investigation of various limits from the perspective of quantum group contractions \cite{CGST2}. (We remark that in cases C, D and G the quantisation in the initial DD basis \eqref{agd} is known~\cite{ab1}).
As the classical DD structures and their duals  can be viewed as the first-oder approximation of the associated quantum groups and non-commutative spaces,
it would be interesting to construct the full quantum models ---in all orders in the deformation parameter--- and to compare their implications with other non-commutative models for quantum gravity.


\section*{Acknowledgments}

This work was partially supported by the Spanish MICINN under grant  MTM2010-18556 (with EU-FEDER support), by Junta de Castilla y Le\'on (A.B. mobility grant EDU/1278/2011) and by the DFG Emmy-Noether fellowship ME 3425/1-1.


\end{document}